# Emerging chirality and moiré dynamics in twisted layered material heterostructures

Andrea Silva,[1,2,*] Xiang Gao,[3,4,*] Melisa M. Gianetti,[5,6] Roberto Guerra,[5] Nicola Manini,[5] Andrea Vanossi,[1,2,#] Michael Urbakh,[3,#] Oded Hod[3]

[1] CNR-IOM – Istituto Officina dei Materiali, c/o SISSA, Via Bonomea 265, 34136, Trieste, Italy
[2] International School for Advanced Studies (SISSA), Via Bonomea 265, 34136, Trieste, Italy
[3] School of Chemistry and The Sackler Center for Computational Molecular and Materials Science, Tel Aviv University, Tel Aviv 6997801, Israel
[4] Department of Modern Mechanics, University of Science and Technology of China, Hefei, Anhui 230026, China
[5] Dipartimento di Fisica, Università degli Studi di Milano, Via Celoria 16, 20133 Milano, Italy
[6] Institutt for maskinteknikk og produksjon, NTNU, Richard Birkelands vei 2B, 7034 Trondheim, Norway

[*] These authors contributed equally
[#] Corresponding authors: vanossi@sissa.it, urbakh@tauex.tau.ac.il

## Abstract

Moiré superstructures arising at twisted 2D interfaces have recently attracted the attention of the scientific community due to exotic quantum states and unique mechanical and tribological behaviors that they exhibit. Here, we predict the emergence of chiral distortions in twisted layered interfaces of finite dimensions. This phenomenon originates in intricate interplay between interfacial interactions and contact boundary constraints. A metric termed the fractional chiral area, is introduced to quantify the overall chirality of the moiré superstructure and to characterize its spatial distribution. Despite the equilibrium nature of the discovered energetic and structural chirality effects they are shown to be manifested in the twisting dynamics of layered interfaces, which demonstrates a continuous transition from stick-slip to smooth rotation with no external trigger.

Recent years saw a surge of interest in 2D materials[1,2,3,4] and, in particular, the fascinating effects of a relative angular rotation (or twist) between their layers.[5] Starting from the discovery of unconventional superconductivity in magic-angle twisted bilayer graphene,[6,7] the equilibrium physical properties of layered materials have been widely investigated. This signified the birth of the field of "twistronic", where electronic tuneability is achieved via twist-angle dependent moiré superstructure variations.[8-17]

The out-of-equilibrium twisting dynamics of these architectures has recently raised interest, as well. Theoretical studies[14,20-25] have investigated the interplay between twist angle, moiré energetics, and strain localization to control the mechanical response of nanoscale van der Waals interfaces.[18-17] This was followed by the development of state-of-the-art experimental techniques that allowed for the precise control over the exerted torque and imposed twist angle.[26-28] Such tools, led to the discovery of a moiré-mediated coupling between translational and rotational degrees of freedom.[27]

Finite-sized contacts introduce an additional factor, namely, edge constraints. Their interplay with surface elasticity and interfacial interactions may induce phenomena that cannot take place in extended, nominally infinite, interfaces. Specifically, the different nature of the edge and surface interactions of a finite 2D material flake with the underlying substrate and the corresponding lattice commensurability conditions can lead to the accumulation of elastic stresses that depend on the twist angle. Here, we demonstrate that such stresses can break inversion symmetry and lead to chiral structural distortions. These, in turn, are manifested in the contact twisting dynamics that demonstrates a continuous transition from stick-slip to smooth rotation.

## Results and Discussion

Our model system consists of a finite-sized heterogeneous contact formed between a circular Bernal-stacked tri-layer graphene flake and an extended AA'A stacked tri-layer *h*-BN substrate (see Fig. 1a, b and coordinate file in the Supplementary Information). Periodic boundary conditions (PBCs) are applied in the lateral directions for the *h*-BN substrate with super-cell dimensions of 37.3 nm x 37.3 nm, assuring that inter-flake interactions are negligible. The dangling bonds at the edges of the graphitic flake are passivated by hydrogen atoms.[29,30] Intralayer interactions in graphene and *h*-BN are described using the REBO[31] and modified Tersoff[24,32] classical force-fields, respectively, whereas the interlayer interactions at the homogeneous and heterogeneous interfaces are described using dedicated anisotropic interlayer potentials (ILP).[24,33,34] The outermost *h*-BN and graphene layers are kept rigid, to mimic the effect of the supporting substrate and a rotating stage, respectively. This setup allows us to impose a well-defined twist angle and measure unambiguously the resulting torque. Langevin damping is applied to the middle layer of each tri-layer countersurface to mimic energy dissipation into the bulk. This setup assures that the twisted heterogeneous inner contact follows conservative dynamics, while dissipating any excess energy (see Methods section for more details).

In order to study the interfacial twist dynamics, we performed two types of simulations: (i) quasi-static calculations, where after each twist step of 0.02° imposed to the top rigid graphene layer the system is allowed to fully relax, thus mimicking slow rotation (see Methods section for details regarding the minimization procedure),[35,36,37,38] and (ii) finite angular velocity simulations, where the top rigid graphene layer is rotated with respect to the bottom rigid *h*-BN layer at a "twisting" angular velocity of $\omega$=0.02°/ps. The torque $\tau$ is defined as the negative of that experienced by the rigid rotating stage due to interaction with the underlying layers. We note that at the limit $\omega \rightarrow 0$, the two simulation protocols give very close results, as expected (see SI Fig. S1). In the dynamical case, we also probe the response of softer contacts, where the top graphene layer is attached via springs with varying stiffness to a rigid rotating stage. This effectively complements the inter-layer interactions within the graphitic flake with springs of controllable stiffness, thus allowing to mimic the presence of an extended contact.

Due to the 1.8% lattice vector mismatch between graphene and *h*-BN lattices, moiré patterns emerge already at the aligned heterogeneous interface.[14] These patterns strongly depend on the twist angle, as shown in Fig. 1c-f. To quantify the corresponding local tensile/compressive strain, for each carbon atom in the bottom graphene layer we compute the average bond-length deviation from the equilibrium value $\delta_i = \sum_{j=1}^{3} \frac{(d_{ij}-d_{cc}^0)}{(3d_{cc}^0)}$, where $d_{ij}$ is the distance between carbon atom $i$ and its $j^{\text{th}}$ nearest-neighbor, and $d_{cc}^0 = 142 \text{ pm}$ is the equilibrium carbon-carbon distance. At the center of the moiré tiles the covalent inter-carbon bonds expand (red regions of positive strain in Fig. 1 c-f) to enhance local commensuration with the underlying *h*-BN layer that possesses a nearest-neighbor equilibrium distance of $d_{\text{BN}}^0 = 145 \text{ pm}$. Correspondingly, regions near the edges of the moiré tiles contract (negative-strain blue regions in Fig. 1c-f). Since the elastic contact has finite lateral dimension, the moiré superstructure deviates from the periodic pattern in the vicinity of its rim. With increasing twist angle $\theta$, the moiré supercell spacing $\lambda(\theta)$, decreases, resulting in the emergence of new moiré tiles at the circumference of the flake. Notably, at certain twist angles (depending on the contact size and mismatch) these new tiles distort, leading to the formation of chiral patterns in the relative deformation map, i.e. the local tensile/compressive strain (see Fig. 1d, f). Hence, the quasi-static twist trajectory is characterized by an alternation between chiral and achiral deformation patterns.

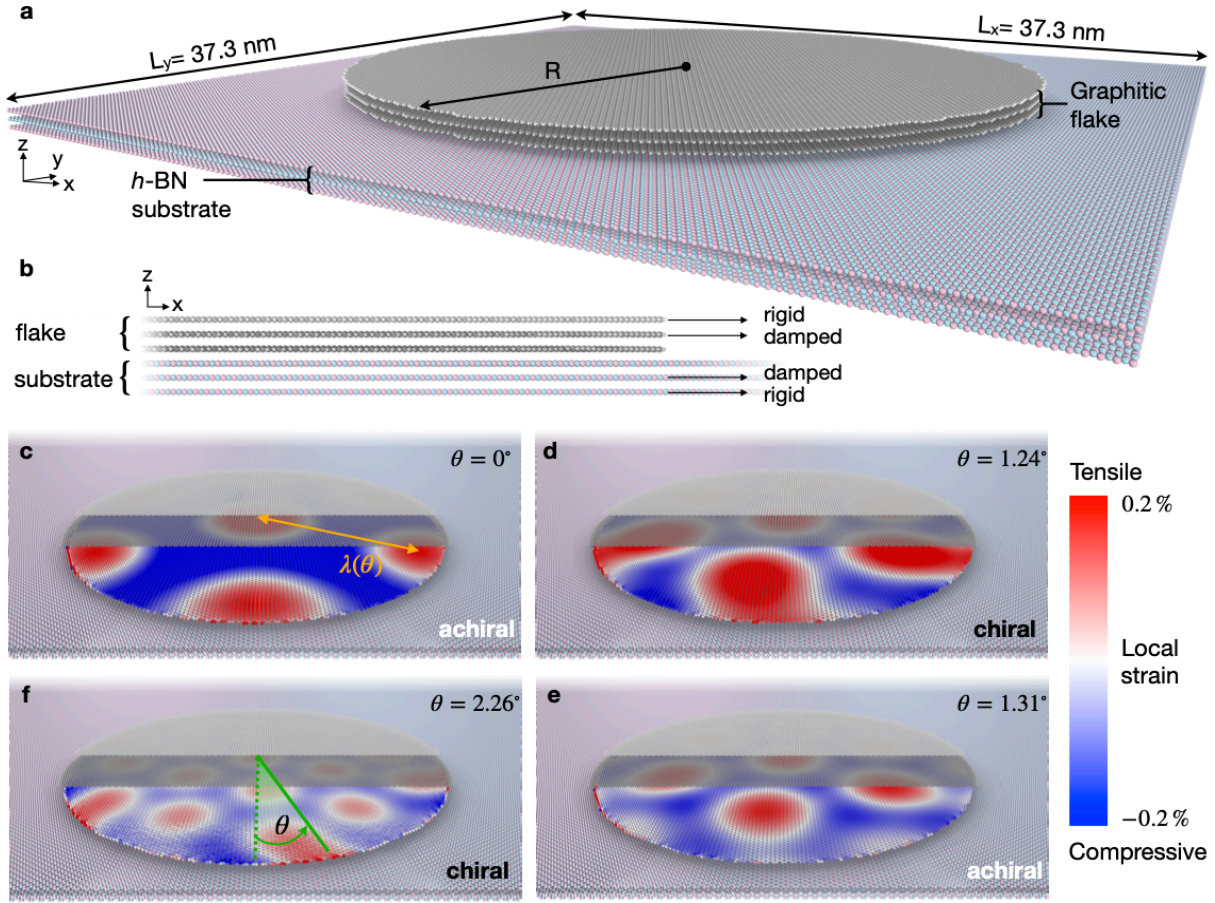

**Fig. 1: Emergence of chirality.** Perspective (a) and side (b) views of the model system, consisting of a circular three-layer graphitic flake of radius $R$ in contact with an extended $h$-BN substrate. (c-f) Relative deformation of the interfacial graphene layer at different angles, as indicated in the top right corner of each panel. The two top layers are depicted partially to expose the interfacial graphene layer, which is colored according to the local strain (see color bar). The direction of increasing twist rotation $\theta$ (counterclockwise) is sketched in panel f (not to scale). At $\theta = 0°$ (c) the interfacial graphene layer deforms in an achiral pattern following the moiré lattice, with tensile strain at the center of each moiré tile and compressive strain at the tile edges. The moiré supercell size $\lambda$ (orange arrow in panel c) decreases as $\theta$ grows (up to 30°). As a result, the interface graphene layer acquires a chiral deformation pattern (d). As subsequent moiré tiles enter through the flake edge, the relative deformation alternates between achiral (e) and chiral (f) patterns.

To quantify this chiral/achiral transition and its relation to the physical properties of the interface, we introduce the concept of fractional chiral area (F$CA$), measuring which fraction of the deformation map (e.g., Fig. 1c-f) cannot be superimposed on its mirror image, and is hence chiral. The F$CA$ is defined as follows:

$$FCA_\xi(\theta) = \min_\alpha \frac{1}{A} \int_A d\vec{r}\, \Theta\left(\left|P_\theta(\vec{r}) - M_\theta(\vec{r};\alpha)\right| - \xi\right), \quad (1)$$

where $P_\theta(\vec{r})$ is the map of relative deformation ($\delta_i$) at a twist angle $\theta$, interpolated over a regular Cartesian 2D grid (with grid spacing of 45 pm), $M_\theta(\vec{r}; \alpha)$ is the original map $P_\theta(\vec{r})$ mirrored along the $yz$ plane crossing the center of the flake and then rotated (around the $z$ axis) by an angle $\alpha$ relative to the $x$ axis, $A$ is the surface area, $A = \pi R^2$, of the circular graphene flake of radius $R$, and $\Theta$ is the Heaviside step function. The threshold parameter $\xi$ controls the sensitivity of the *FCA* function to the difference between the pattern and its mirror image. The minimization is performed over the angle $\alpha$ searching for the best possible superposition between the deformation map and its mirror image. With this definition, the fractional chiral area is bound in the range $0 \leq FCA_\xi(\theta) \leq 1$, where $FCA_\xi(\theta) = 0$ is obtained when the deformation map is achiral (perfect overlap between $P_\theta$ and $M_\theta$) and $FCA_\xi(\theta) = 1$ when the entire surface area is chiral. In practice, for any given twist angle $\theta$, we consider the deformation map of the interfacial graphene layer, interpolate it on a 2D Cartesian grid, construct its mirror image, such that its center coincides with that of the original map, rotate the mirror image map around the $z$ axis crossing the center by an angle $\alpha$, and evaluate the fractional chiral area, by searching for a minimum of the nonmatching area fraction as a function of $\alpha$; see Section 5 of the SI for further details. We note that an alternative chirality measure was provided for rigid bodies, based on the Hausdorff distance.[39,40,41] In SI section 5 we generalize this definition using the Pearson cross-correlation function, yielding a continuous chirality index (*CI*). While the *CI* and *FCA* correlate well (see SI section 5 and SI Fig. S7), the latter metric enables comparison between deformation maps of different amplitudes as it is based on total area normalization rather than on deformation map amplitude normalization.

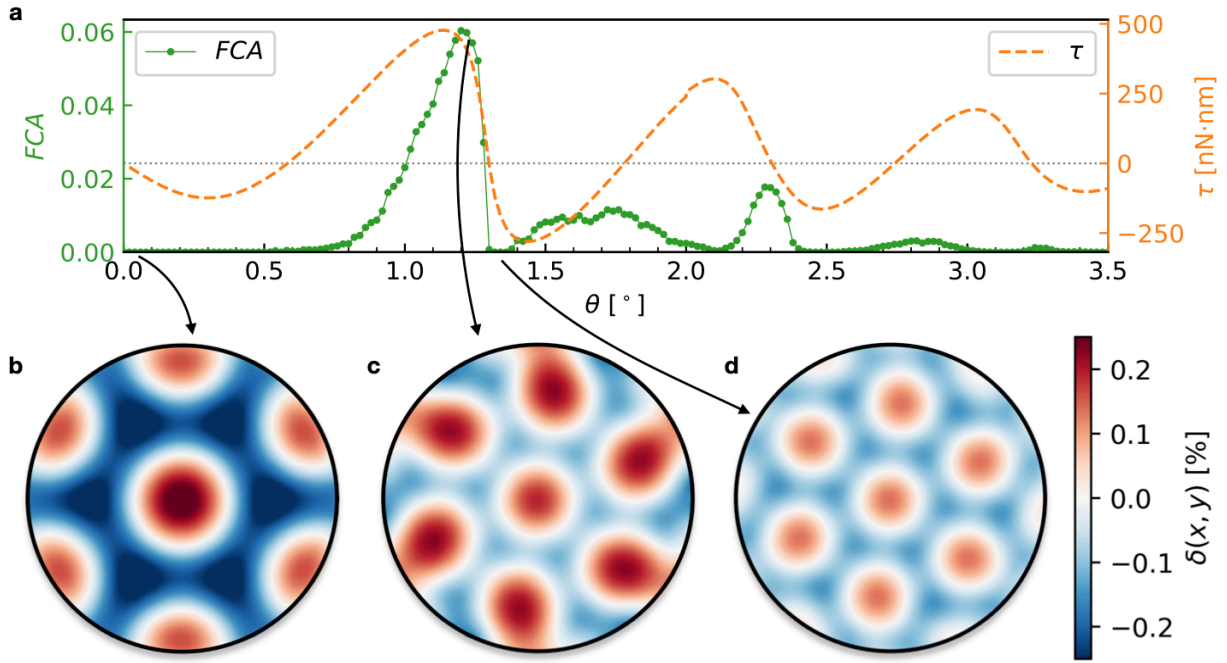

**Fig. 2: Chirality fingerprint in the relative deformation map of an $R = 14$ nm graphene/h-BN interface.** Quasistatic twisting leads to elastic deformations of the interface, often characterized by a chiral pattern: (a) fractional chiral area $FCA_{\xi=0.001}(\theta)$ (solid green, left axis) and torque (dashed orange, right axis) as a function of the twist angle. In the lower panels, the strain patterns correspond to (b) the aligned configuration, (c) an $FCA$ maximum and (d) a twisted $FCA$ minimum. The color code expresses the local strain according to the color bar at the right.

Fig. 2a compares $FCA_{\xi=0.001}$ with the torque generated by the interaction with the h-BN substrate, as a function of twist angle, from quasistatic twisting simulations. Following the initial aligned configuration, where the $FCA$ vanishes (see the corresponding pattern in Fig. 2b), the $FCA$ displays a sequence of peaks, e.g. at $\theta = 1.24°$ (see Fig. 2c) and $\theta = 2.26°$. Notably the $FCA$ peaks appear close to the angle of maximum torque variation (dashed orange line in Fig. 2a), correlating with the maxima and minima of interfacial angular stiffness, $\partial\tau/\partial\theta$, which occur when the system deforms to oppose rotation or when it jumps towards the next angular energy minimum, respectively (see SI Fig. S14). A similar effect is also found for a homogeneous graphitic interface between a circular tri-layer graphene flake and an extended tri-layer graphite substrate model (see SI section 8). This finding rules out the heterogeneous nature of the system depicted in Fig. 1 as the cause of the observed chirality.

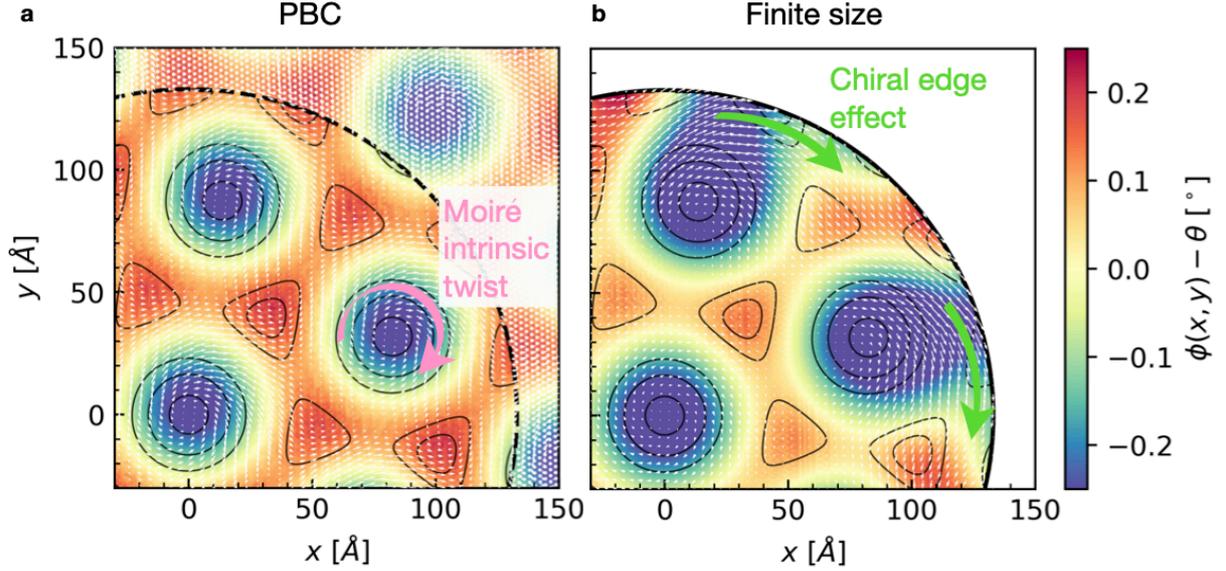

**Fig. 3: The origin of twisting chirality.** Local twist angle (see Methods section) in a periodic (a) and finite-size (b) systems rotated by the same twist angle $\theta = 1.24°$. The color scale indicates the deviation between the local twist angle $\phi$ and the imposed global twist angle $\theta$. White arrows report the local displacement of the atoms relative to a rigidly rotated crystal, i.e. the displacement field, magnified by a factor 50 and 22 for panels (a) and (b), respectively. For better visualization, the local angle and displacement vector field are interpolated in the two panels over grids of different densities (700x700 and 140x140 points, respectively). For the laterally periodic system of panel (a) the black arc of radius $R = 14$ nm marks the region corresponding to the finite flake in panel (b); the region outside this circle is made partially transparent to obtain better contrast. As a guide to the eye, the black contours highlight the interlayer energy landscape (circles mark minima positions and triangle maxima positions) according to an analytical stacking energy model (see Methods section). The pink arrow in panel (a) marks the handedness of the intrinsic moiré twist and the green arrows in panel (b) indicate the direction of the chiral deformation.

The question then arises: what is the origin of the observed twisting chirality? To address this, we plot in Fig. 3 the angular deformation map, defined as the angular bond distortion relative to the aligned stacking configuration around each atomic position for infinite and finite circular heterogeneous graphene/*h*-BN interfaces (see Methods section for further details). A hint toward the origin of the emerging chiral deformation patterns is given by the fact that they are absent in extended twisted interfaces (see Fig. 3a). This suggests that the edge of the circular flake plays a leading role in the formation of the chiral patterns.

Focusing first on the *infinite* aligned ($\theta = 0°$) system, following structural relaxation, local compressive and tensile strains arise due to the tendency of the graphene layer to comply with the underlying *h*-BN substrate lattice (see Fig. S2).[42] At a finite twist angle, the graphene layer is forced away from its optimal stacking configuration above the *h*-BN substrate, thus moiré patches develop vortex-type lateral atomic displacements to improve the local stacking (see white arrows in Fig. 3a).[14,43,42,44,45] Notably, in the infinite contact, the resulting angular deformation maps break neither mirror nor inversion symmetries.

Near the center of the *finite* circular flake, the angular deformation map resembles that of the infinite contact. In contrast, near the flake edge, the enhanced flexibility allows the atoms to comply better with the underlying substrate lattice. Except for those specific twist angles, where the circle encompasses full moiré tiles (see for example Fig. 2d), this results in a non-radial expansion of the vortex displacement vector field toward the edge (see white vector map in Fig. 3b). In turn, chiral "fan-like" structures emerge in the angular deformation map (see colormap in Fig. 3b). The reconstructed edge includes regions of enhanced stacking commensurability that pin to their position as the twist angle is changed. This pinning persists up to the point where a complete moiré tile is encompassed by the circle, where a sudden jump in the angular deformation map, from chiral to achiral pattern, occurs (see supplementary movie 1). Starting from the $\theta = 0°$ configuration and reversing the twist direction results in a mirror-image trajectory (see Fig. S3).

We note that the mechanism underlying the chiral twist distortion is robust against contact size, thickness, shape, initial aligned stacking configuration, and chemical composition. As shown in SI Fig. S8 a larger circular graphene flake of radius $R = 21$ nm exhibits clear fingerprints of chirality in the strain map when twisted in contact with the $h$-BN substrate. Furthermore, reducing the circular flake thickness to a single layer does not eliminate the observed chiral patterns (see Fig. S9, S10). As this mechanism takes place at the moiré-tile scale, we expect it to be largely independent of the shape of the edge: Fig. S17 reports the strain map of a hexagonal, rather than circular, graphitic flake, which exhibits a chiral/achiral alternation during twisting. Finally, as mentioned above, Figs. S12-S13 demonstrate the same phenomenon for a twisted homogeneous graphitic interface regardless of the adopted aligned stacking configuration, AB or AA.

While originating in edge stacking energy considerations, the chiral deformation phenomenon discovered herein should have manifestation also in the twisting dynamics of finite layered interfaces. To explore this, we performed molecular dynamics simulations at finite angular velocity. To that end, we attached the top rigid graphene layer to a rigid stage through a set of lateral springs of constant $K$, as sketched in Fig. 4a. Varying the stiffness of these springs allows us to mimic the twisting elasticity of a graphitic multilayer of variable thickness, ranging from a three-layer setup ($K \to \infty$) to a 10.4 nm thick bulk "pillar" stacked above the uppermost graphene flake (see Methods section for further simulation details).

Fig. 4b reports the counterclockwise (CCW) and clockwise (CW) torque traces obtained for an $R = 14$ nm circular graphene flake twisted at an angular velocity of $\omega = 0.02°/\text{ps}$ above an $h$-BN substrate. The whole rotation protocol comprises a CCW trace from $0°$ to $20°$ and a back trace from $20°$ back to $0°$. The dynamics exhibits rotational stick-slip motion,[46] manifesting itself as hysteresis regions (grey areas) between the torque trace and retrace, linked to the interplay between intralayer elasticity and interfacial edge commensurability and pinning conditions. The dynamic slip events are associated with chirality jumps in the deformation maps (see SI Fig. S15) and occur at the same angles as in the quasi-static simulations. Unlike its shear-sliding counterpart,[47,48,49,50] the twisting dynamics exhibits a transition from stick-slip to smooth rotation as $\theta$ increases from 0° to 30°, due to the progressive reduction in moiré-tile size and out-of-plane corrugation, and the resulting edge pinning weakening.

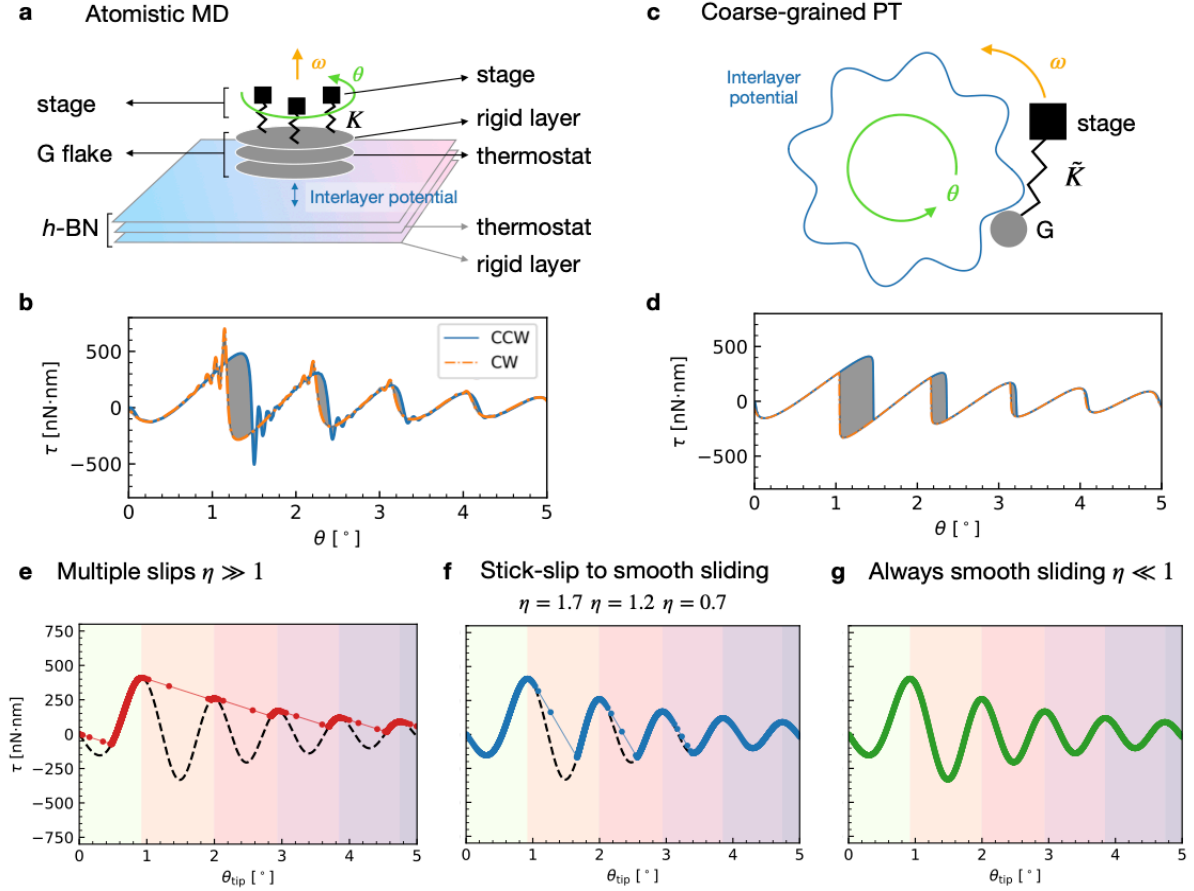

**Fig. 4: Dissipative rotational dynamics.** (a) Sketch of the atomistic MD-simulation model showing the direction of constant angular velocity $\omega = 0.02$ °/ps imposed on the stage. The stage is connected to the graphene top (rigid) layer via six springs of constant $K = 500$ N/m, anchored $R_s = 7$ nm off the center of the flake. (b) MD simulation results of the torque as a function of angular stage position, $\theta$, for the counterclockwise (blue solid curve) and clockwise (orange dot-dashed curve) parts of the rotation loop. (c) Effective Prandtl-Tomlinson (PT) model describing the system: the top layer rotating at constant angular velocity is mapped to a virtual stage moving at the same angular velocity. The stage is connected to a point-like object (G), which represents the graphitic flake that is rotated through an angular spring with constant $\widetilde{K} = 1.2$ $\mu$N·nm/deg$^2$. (d) Effective PT friction loop (same line convention as in panel b). The hysteresis is highlighted as a shaded area. (e)-(g) Effective PT model results of the torque (colored lines) as a function of tip angular position, $\theta_{\text{tip}}$, for driving spring stiffnesses of (e) $\widetilde{K} = 0.2$, (f) $\widetilde{K} = 1.2$, and (g) $\widetilde{K} = 10$ $\mu$N·nm/deg$^2$. Black dashed lines represent the angular derivative of the PT substrate potential. Background coloring indicates regions of different effective PT parameter $\eta$ (see text for further details).

This complex behavior can be understood and characterized quantitatively in terms of a generalized Prandtl-Tomlinson (PT) model.[51,52,53,54] The system presented in Fig. 4a is coarse-grained into a one-dimensional model (see Fig. 4c), where the graphitic stack is represented by a point-like object attached via an effective rotational spring $\widetilde{K}$ to a stage moving at constant angular velocity, $\omega$. The slider rotates over the corrugation generated by the interaction with the underlying h-BN substrate, here modelled by a fixed angular potential (see Methods section).[27,55] The torque trace of the atomistic simulation (Fig. 4b) is well reproduced by the coarse-grained model (Fig. 4d). The first three peaks are characterized by a clear hysteresis between the CCW and CW loops due to the stick-slip dynamics, while for increasing misalignment the rotation proceeds smoothly. We note that the high frequency oscillations apparent in the MD simulations are absent in our PT model results as we deliberately run the latter at much lower angular velocities using higher damping to approach experimentally accessible conditions.

To understand the mechanism that drives the intrinsic transition from stick-slip to smooth sliding rotation with increasing twist angle, we first recall the standard PT model for sliding motion. For the standard periodic potential of the form $U(x) = V_0 \cos(2\pi x/a)$, the regime of motion is determined by the dimensionless parameter $\eta = \max[d^2 U(x)/dx^2]/K_{\text{PT}} = 4\pi^2 V_0/(K_{\text{PT}} a^2)$, such that stick-slip occurs for $\eta > 1$ and otherwise smooth sliding is obtained. Here, $a$ is the period of the potential and $K_{\text{PT}}$ is the stiffness of the driving spring. In that case, $\eta$ is constant (and so is the sliding regime) throughout the dynamics. Conversely, for rotational motion, the potential amplitude, $V_0$, becomes $\theta$-dependent, due to the moiré pattern evolution. Like the standard PT model, in the rotational case one can still tune the stiffness of the external spring, to obtain strictly (multiple) stick-slip ($\eta \gg 1$, Fig. 4e) or smooth sliding ($\eta < 1$, Fig. 4g). However, for a broad range of spring constants a transition from stick-slip motion ($\eta > 1$) to smooth sliding ($\eta < 1$) occurs during the twist dynamics (Fig. 4f). The different regimes are dictated by the following expression, $\eta_I = \max_I [d^2 U(\theta_{\text{tip}})/d\theta_{\text{tip}}^2]/\widetilde{K}$, where $\theta_{\text{tip}}$ is the instantaneous slider angular position, distinct from the imposed rotating stage angle $\theta(t) = \omega t$. The letter $I$ signifies that in the rotational case the maximum is evaluated over a given angular interval between successive local maxima of the torque, as indicated by the color shadings in Fig. 4e-g. $\eta_I$ is the local analogue of the standard PT parameter for the case of rotational moiré motion.

In practice, the effective rotational spring constant (and hence the sliding regime) is controlled by the stack thickness. Approximating the graphitic stack as a continuum body, the rotational stiffness can be expressed as $\widetilde{K} = GJ/L$,[56] where $G \approx 2$ GPa is the $c$-axis shear modulus of graphite,[57,58] $J = \pi R^4/2$ is the torsional constant[56] of a circular stack of radius $R$, and $L$ is the stack thickness. From these relations we can estimate the effective stack thickness corresponding to the simulation setup used to obtain the results presented in Fig. 4b. Given six springs of $K = 500$ N/m, which connect the rotating stage to the top layer of the graphitic stack at a distance $R_s = 7$ nm from the center and angularly separated by $60°$ (see Fig. S16), the effective rotational stiffness is $\widetilde{K} = 6KR_s^2 \approx 2.6\ \mu\text{N} \cdot \text{nm}/\text{deg}^2$. This, in turn, corresponds to an effective thickness of:

$$L = \frac{GJ}{6KR_S^2} = 0.8 \text{ nm} \approx 3 \text{ layers}. \qquad (2)$$

Hence, the results of Fig. 4a correspond to a total 6-layered graphitic stack rotating above an *h*-BN substrate (see Fig. S4). While a slab of such thickness is still feasible to simulate fully atomistically, thicker graphitic slabs would require substantial computational efforts, which can be avoided by adopting softer spring constants when driving the 3-layer atomistic model. Alternatively, if fine details of the torque trace are less important, one could use the coarse-grained PT model. We note that the qualitative nature of the rotational moiré stick-slip motion is exhibited even for a single graphene layer rotating above and *h*-BN stack, as long as the edges are left flexible (see SI Section 7 and Fig. S11).

## Conclusions

Twist dynamics is emerging as a means to control the structural, mechanical, tribological, and electronic properties of layered contacts.[28,59,60] While one may try to naively extrapolate sliding phenomena to understand twist dynamics, our findings predict that the latter exhibits an intriguing complex behavior originating from the intricate interplay between interlayer elasticity, inter-lattice commensurability, and twist-dependent edge pinning. In structurally superlubric interfaces, edge energy dissipation and pinning effects dominate friction[37,61,62]. Hence, edge deformation can have important consequences on the overall dynamical behavior of the system. Most notably, we find an intrinsic transition from stick-slip to smooth rotation, accompanied by transitions from chiral to achiral deformation maps. We note that Yang and coworkers[46] also identified a stick-slip to smooth rotation transition, controlled via an external parameter, namely the substrate strain. This is fundamentally different from our prediction, where the transition emerges due to the intrinsic changes in contact geometry and interactions during rotation, without any external control. The effects predicted here could be observed in rotating islands of magnetic colloids[63,64] over an appropriate substrate and in nanoscale contacts with recently developed experimental setups.[28,65,66,67] Beyond their fundamental-science value, such effects should be taken into consideration when designing twisting layered-material contacts.[68] The graphitic systems considered harbour localized electronic states, similar to the zigzag edge states that carry a unique spin structure.[69,70,71,72,73] Such electronic states may be affected by structural distortions of the ionic degree of freedom, inheriting the chiral nature of the latter.

## Methods

**MD simulation details**

We adopt LAMMPS[74] as our simulation platform. Our graphene/*h*-BN heterostructure model system consists of a Bernal (ABA) stacked three-layer-thick circular graphitic flake saturated by hydrogen atoms, residing atop a three-layer periodic AA'A stacked *h*-BN substrate. The center of the middle hexagon ring of the bottom graphene layer is placed atop a nitrogen atom, forming the lowest energetic local stacking. In this initial aligned configuration, the interface exhibits the largest moiré superstructure of period of $\lambda(0°) \approx 14$ nm. Periodic boundary conditions are applied along both lateral directions of the *h*-BN substrate, leaving 10 nm between periodic images of the graphitic flake to avoid spurious interactions. To mimic a flat stationary support, the bottom *h*-BN layer is kept fixed at its initial position.

The intralayer interactions are described by the REBO potential[31] for graphene and the shifted Tersoff potential[32,24] for *h*-BN. The interlayer interactions are accounted for by the registry-dependent ILP.[23,24,33,34]

Two types of dynamic atomistic simulations are performed. In the first, the twist angle is imposed directly to the rigid top graphene layer, and the two lower-layers of the graphitic flake and two upper *h*-BN layers are allowed to respond and relax according to the force field. To model thicker flakes, a second type of dynamic simulation is performed, where the top rigid graphene layer is connected by lateral springs to 6 dummy atoms initially located atop six carbon atoms equally angularly spaced at a distance $R_s = R/2 = 7$ nm, and the graphitic flake is driven by rotating the dummy atoms. In both models, the imposed twist angle is varied at a constant angular velocity of $\omega = 0.02°/\text{ps}$. Viscous damping is applied to the middle layers of the graphene and *h*-BN stacks to dissipate the excess energy. The external torque is calculated and recorded by evaluating the opposite of the instantaneous torque experienced by the top rigid layer or by the dummy atoms at any given imposed twist angle, for the two simulation schemes, respectively.

In addition to these dynamic simulations, quasistatic simulations are also performed to study the slow rotation limit. To that end, the top rigid graphene layer is rotated in steps of $0.02°$, allowing the entire system (apart from the fixed support) to relax after each step using first the conjugate gradient algorithm with an energy tolerance of $10^{-15}$ eV, and then the FIRE algorithm[75] with a force tolerance of $10^{-4}$ eV/Å.

As baseline we also performed fully rigid calculations, where a rigid circular single-layer graphene flake was placed atop a periodic rigid single-layer *h*-BN substrate at an interlayer distance of 3.34 Å in the same initial stacking as for the dynamic simulations. The graphene layer was rotated in steps of 0.005°, while computing the energy and torque at each twist angle.

**Local angle**

The local twist angle is defined at any given atomic position as the nearest-neighbour bond angle relative to the *x*-axis (chosen to be parallel to armchair direction of the fixed *h*-BN support layer) modulo 60° and averaged over all nearest neighbors.

**Rigid model**

With the use of the atomistic rigid model calculations, we developed continuum analytical expressions for the interfacial energy and torque to be used in the Prandtl-Tomlinson calculations. To that end, the periodic interlayer potential energy per unit area was described in terms of the evolution of the moiré pattern with the twist angle, $\theta$, as follows:[55,76]

$$W(x,y,\theta) = U_1 - \frac{2}{9} U_0 \left( 2\cos\frac{2\pi x}{\sqrt{3}\lambda(\theta)} \cos\frac{2\pi y}{\lambda(\theta)} + \cos\frac{4\pi x}{\sqrt{3}\lambda(\theta)} \right), \quad (3)$$

where $U_0 = 0.003$ eV/Å² and $U_1 = -0.017$ eV/Å² are the energy density parameters, that we fit against the potential energy profile calculated using the ILP, and $\lambda(\theta)$ is the twist-angle dependent period of the moiré supercell given by:[14,76]

$$\lambda(\theta) = \frac{\sqrt{3}d_{CC}^0 \rho}{\sqrt{2\rho(1-\cos\theta)+(\rho-1)^2}}, \quad (4)$$

where $d_{CC}^0 = 142.039$ pm is the equilibrium carbon-carbon distance of graphene, $\rho = \frac{d_{BN}^0}{d_{CC}^0}$ is the mismatch ($d_{BN}^0 = 145.596$ pm) and $\theta$ is the misalignment angle. We note that a different choice of initial stacking in the atomistic simulations would require a modified potential.

To represent a circular flake, we used a circular portion of radius $R \propto N^{0.5}$ out of the periodic potential (Eq. 3), with $N$ being the number of atoms in contact. Due to the circular symmetry of the flake, the misfit between the moiré lattice orientation and the twist angle plays no role. The resulting approximate energy of the whole flake as a function of twist angle is obtained by integrating $W$ over the circular area[27,55], yielding:

$$U(0,0,\theta) = U_1 \pi R^2 - \frac{\sqrt{3}U_0}{3} \lambda(\theta) R \, J_1\left(\frac{4\pi}{\sqrt{3}\lambda(\theta)} R\right), \quad (5)$$

where $J_1(x)$ is the first-order Bessel function of the first kind. By differentiating with respect to the misalignment angle $\theta$, we obtain the torque acting on the flake:

$$\tau(0,0,\theta) = \frac{\partial U}{\partial \theta} = U_0 \pi \frac{4}{9} \frac{1}{d_{CC}^0 d_{BN}^0} \sin(\theta) (\lambda(\theta)R)^2 J_2\left(\frac{4\pi}{\sqrt{3}\lambda(\theta)} R\right). \tag{6}$$

Given this expression, we see that the oscillatory torque behavior, observed by the dashed lines in Fig. 4e-g, reflects the evolution of the moiré period with twist angle. As $\theta$ increases $\lambda(\theta)$ decreases, resulting in new moiré tiles entering the flake area, as manifested by the $J_2$ oscillations.

**PT model details**

The analytic potential derived above was then used in the modified PT model presented in the results section. The following equation of motion was propagated:

$$I\ddot{\theta}_{\text{tip}} = -\frac{\partial U}{\partial \theta_{\text{tip}}} - \widetilde{K}(\theta_{\text{tip}} - \omega t) - I\gamma \dot{\theta}_{\text{tip}} \tag{7}$$

using the fifth-order Runge-Kutta algorithm[78,79] with a fixed time-step of $0.005$ ps. Here, $I = 0.8 \times 10^8$ g/mol nm$^2$ is the moment of inertia of the three-layer flake of radius $R = 14$ nm and $\gamma = 166$ ps$^{-1}$ is the damping parameter. The stage was rotated at a constant angular speed of $\omega = 2.45 \times 10^{-5}$ °/ps. We note that the relation between $\widetilde{K}$ and $K$ was discussed above.

**Supporting Information** Benchmark of the simulation protocol and consistency with the analytical models, details of the chirality metrics, robustness of the chirality effect against contact size, chemical composition, number of layers, and shape.

**Acknowledgment**

A.S., M.G., N.M., and A.V. acknowledge support from the grant PRIN2017 UTFROM of the Italian Ministry of University and Research. X. G. acknowledges the postdoctoral fellowships of the Sackler Center for Computational Molecular and Materials Science and the Ratner Center for Single Molecule Science at Tel Aviv University. A.S. and A. V. acknowledges also support by ERC Advanced Grant ULTRA-DISS, contract no. 8344023. O.H. is grateful for the financial support from the Israel Science Foundation (grant no. 1586/17), Tel Aviv University Center for Nanoscience and Nanotechnology, the Naomi Foundation by the 2017 Kadar Award and the Heineman Chair of Physical Chemistry. M.U. acknowledges the financial support from the Israel Science Foundation (grant no. 1141/18) and the ISF-NSFC joint grant 3191/19.